\documentclass[aa,superscriptaddress,amsfonts,amssymb,amsmath,showpacs,twocolumn,nofootinbib]{revtex4-2}
\usepackage{bm}
\usepackage{amsfonts}
\usepackage{latexsym}
\usepackage{graphicx}
\usepackage{amsmath}
\usepackage{cancel} 
\usepackage{soul}
\usepackage{palatino}
\usepackage{xcolor} 
\usepackage{mathpazo}
\usepackage{tensor}
\usepackage{textcomp}
\usepackage{float}
\usepackage{booktabs}
\usepackage{dcolumn}
\usepackage{booktabs}
\usepackage{multirow}
\usepackage{hyperref}
\hypersetup{colorlinks,citecolor=blue}
\usepackage{amsmath}
\usepackage{xcolor}
\usepackage{orcidlink}
\usepackage{epsfig}
\usepackage{caption}
\usepackage{subcaption}
\usepackage{commath}
\hypersetup{colorlinks,citecolor=blue}
\hypersetup{colorlinks=true,linkcolor=magenta,filecolor=magenta,    urlcolor=blue}

\def\be{\begin{equation}}
\def\ee{\end{equation}}
\def\bea{\begin{eqnarray}}
\def\eea{\end{eqnarray}}

\begin{document}

\title{Probing Generalized Emergent Dark Energy with DESI DR2}

\author{Vipin Kumar Sharma}
\email{vipinkumar.sharma@iiap.res.in}
\affiliation{Indian Institute of Astrophysics, Koramangala II Block, Bangalore 560034, India}
\author{Himanshu Chaudhary$^{*}$}
\email{himanshu.chaudhary@ubbcluj.ro,\\
himanshuch1729@gmail.com \textcolor{black}{(Corresponding author)}}
\affiliation{Department of Physics, Babeș-Bolyai University, Kogălniceanu Street, Cluj-Napoca, 400084, Romania}
\affiliation{Research Center of Astrophysics and Cosmology, Khazar University, Baku, AZ1096, 41 Mehseti Street, Azerbaijan}
\author{Sanved Kolekar}
\email{Sanved.kolekar@iiap.res.in}
\affiliation{Indian Institute of Astrophysics, Koramangala II Block, Bangalore 560034, India}

\begin{abstract}
As an update on the initial findings of DESI, the new results provide the first hint of potential deviations from a cosmological constant ($\omega=-1$), which, if confirmed with significance $>(2-4)\sigma$, would challenge the validity of $\Lambda$ within the $\Lambda$CDM model. We explore the Generalized Emergent Dark Energy (GEDE) model using recent BAO measurements from DESI DR2, Type Ia supernova compilations, and CMB distance priors. Employing nested sampling, we constrain the parameter $\Delta$, which characterizes deviations from $\Lambda$CDM. Our analysis shows that with CMB+DESI DR2 alone, GEDE tends to prefer positive values of $\Delta$. However, when different SNe Ia calibrations are included, the model favors negative values of $\Delta$, corresponding to an earlier injection of dark energy. The Marginalized constraints on $\omega(z)$ further shows that GEDE sharply emerges but then asymptotes to $\omega=-1$ without crossing it. At $z \sim 1$ data, GEDE provides a better fit than $\Lambda$CDM, while at $z \lesssim 0.5$ the data favor $\omega > -1$, bringing the model deviate from $\Lambda$CDM. Bayesian model comparison shows weak support for GEDE with CMB+DESI DR2 ($\ln BF=1.96$), moderate with PP ($\ln BF=2.65$), weak-to-moderate with Union3 ($\ln BF=2.34$), and weak with DES-SN5Y ($\ln BF=1.44$). Overall, GEDE is consistent with current data and mildly favored when SNe Ia are included, making it a viable extension of $\Lambda$CDM that merits further investigation with future high precision measurements.
\end{abstract}
\maketitle
%\tableofcontents

\section{Introduction}\label{sec_1}
Following the release of numerous data sets \cite{dark2016dark,alam2021completed,nidever2015data,mellier2024euclid,schlafly2023survey} the conventional $\Lambda$ cold dark matter (CDM) in the present concordance cosmology, which is based on the straightforward six-parameter model, is becoming more noticeable with emerging observational trends now reaching significance levels approaching or exceeding 2-4 $\sigma$. Particle physicists demand 5$\sigma$ to confirm a discovery (see \cite{Lyons:2013yja} for discovering the significance of 5 $\sigma$), cosmologists treat 2–4$\sigma$ deviations as significant, since cosmology depends on unique, unrepeatable observations of the Universe. In the framework of standard $\Lambda$CDM with general relativity background, there are two main ingredients of the Universe; dark matter (DM$\sim 26.5\%$) with zero pressure and  hypothetical dark energy (DE$\sim 68.5\%$) with negative pressure \cite{Planck:2018vyg}. Both the observable, the observed large scale structures and the accelerated expansion of the Universe are caused by these enormous dark sectors without knowing the actual physics of DM and DE.  The phenomenological nature of $\Lambda$CDM has prompted searches for alternative scenarios due to known theoretical problems relating to reconciling the absurdly small value of the comological constant (which is constant in time and behaves like a fluid with fixed energy density)  with the predictions of quantum vacuum theory \citep{Sahni:1999gb,Padmanabhan:2002ji}. Also, the tensions in the Hubble constant $H_0$  and in the amplitude of the growth of structure $\sigma_8$ (alternatively one can also measure the parameter $S_8\equiv\sigma_8\sqrt{\Omega_{m0}/0.3}$) are the signals of the limitations of the $\Lambda$CDM cosmology (see Refs. \citep{Bagla:1995az,Jassal:2004ej,DiValentino:2020zio,Pandey:2019plg,Bhattacharyya:2018fwb,DiValentino:2021izs,Sharma:2023kzr,DiValentino:2020vvd} for overview, and references therein).  Hence, the most vibrant research theme in the last decade focused on attributing the accelerated late-time expansion as an extension of $\Lambda$. In this direction, numerous theoretical proposals have been put forward by researchers, broadly categorized into two physical frameworks for explaining cosmic acceleration: 'Dark Energy (DE)' and 'Modified Gravity' (see \cite{Joyce:2016vqv} for a concise overview). Essentially, within DE models, there are several categories such as Quintom DE model \cite{Caldwell:1997ii,Cai:2025mas,Bayat:2025xfr,Lin:2025gne}, interacting DE models \citep{Chakraborty:2025syu,Kumar:2017dnp,Bjaelde:2012wi,Kumar:2019wfs,Lucca:2020zjb,Yang:2019uog},  Emergent Dark Energy \citep{Hernandez-Almada:2020uyr,Yang:2020ope,Rezaei:2020mrj,Pan:2019hac}, DE models with parameterized EoS (the constant $\omega$ ($\omega$CDM) model, the Chevallier–Polarski–Linder (CPL) model, and the Jassal–Bagla–Padmanabhan (JBP) model) \cite{Chevallier:2000qy,Linder:2002et,Jassal:2004ej} and so on. The Modified Gravity category, for instance, corresponds to $f(R)$ gravity models with different functional forms (including other physical aspects in addition to DE explanation) \cite{Amendola,Tsujikawa:2007xu,Sharma:2022fiw,KumarSharma:2022qdf,Dolgov:2003px,Sharma:2019yix,DeFelice:2010aj,Sharma:2020vex,Sotiriou:2008rp,Sharma:2023vme,afroz2025hint}. It is true, however, that most beyond-$\Lambda$CDM scenarios introduce more free parameters than the six defining a flat $\Lambda$CDM model, which is not good for model‐selection tests. In this context, the Phenomenologically Emergent Dark Energy (PEDE) model \cite{Li:2019yem} is remarkable: it retains exactly six parameters yet resolves the $H_0$ tension at the 1$\sigma$ level and  when confronted with certain combinations of probes is even preferred over $\Lambda$CDM. 

Further, motivated by the appealing possibilities of Emergent DE models \cite{Hernandez-Almada:2020uyr,Yang:2020ope,Rezaei:2020mrj,Pan:2019hac,Li:2019yem}, authors of \cite{Li:2019yem}, have introduced a generalized version of PEDE model \cite{Yang:2021eud}. The PEDE model states that DE had no effective presence in the past and it only emerges in later time in Universe. So they introduced a zero degree of freedom  for DE and the model exhibits a symmetric behavior centered around present epoch, during which the densities of DE and matter are of comparable magnitude. The generalized form called as Generalized Emergent Dark Energy (GEDE) model  \textcolor{red}{\cite{Yang:2021eud,Li:2020ybr}} has a free parameter denoted by $\Delta$ that describes the evolution picture of DE (which was not in PEDE). The other parameter denoted by $z_t$, which is a fixed parameter in the model, describes the transitional redshift where DE density equals to matter density. GEDE model  has the ability to include both PEDE model (for value of $\Delta=1$) and $\Lambda$CDM model (for value of $\Delta=0$) as two of its special cases. This flexibility in the model helps us to understand the behavior of DE evolution with time. This helps to filter out the possibility of misleading results that can be caused by using incorrect form of parameterization of DE evolution. Based on the results \cite{Yang:2021eud,Li:2020ybr}, the GEDE model is worth investigating.

In the present article, we focus on GEDE model and analyse it by considering various observational recent datasets including Baryon Acoustic Oscillation (BAO) data from DE Spectroscopic Instrument Data Release (DESI DR2), Type Ia Supernovae (SNe Ia), and Cosmic Microwave Background (CMB) distance priors. We present a novel methodology for investigating variations in the GEDE parameter ($\Delta$) and conduct a comparative analysis between the GEDE and $\Lambda$CDM models using cosmological tracers that probe late-time, transition-epoch, and early-time DE. The paper is organized as follows. In Section~\ref{sec_2}, we revisit the GEDE model and present its governing equations. Section~\ref{sec_3} describes the methodology employed to constrain the model's free parameters and provides a brief overview of the observational datasets used. The results of our analysis are detailed in Section~\ref{sec_4}. Finally, Section~\ref{sec_5} concludes the paper with a discussion of our findings and their broader implications.

\section{Extension of $\Lambda$ as GEDE}\label{sec_2}
We introduce the GEDE model as an extension of $\Lambda$CDM \cite{Li:2020ybr}. This model suggest that DE component is insignificant in the early Universe but becomes dominant at later times.

For the spatially flat Friedmann-Lematire-Robertson-Walker (FLRW) metric, the evolution of this Universe is governed by the Friedmann equations along with the continuity equations for each component (matter (\(\rho_m\), consisting of both cold dark matter and baryons), radiation (\(\rho_r\)), and GEDE). The first Friedmann equation reads,
\begin{equation}\label{eq_1}
H^2 \equiv \left(\frac{\dot{a}}{a}\right)^2 = \frac{8\pi G}{3} \sum_x\rho_{x},
\end{equation}
The continuity equation in FLRW reads,
\begin{equation}\label{eq_2}
\dot{\rho}_{\text{x}} + 3H(1 + w_{\text{x}}) \rho_{\text{x}} = 0,
\end{equation}

Here, \(H\) is the Hubble parameter, \(a\) is the scale factor, over $(\dot{})$ represents the cosmic time derivative and \(\rho_x\) represents the energy density of each component with x$\in$ (DE, matter, radiation). This integrates to give for constant $\omega$  (-1 for traditional dark energy $\Lambda$, 1/3 for radiation, and zero for dark matter),
\begin{equation}\label{eq_3}
\rho_x=\rho_{x,0}(1+z)^{3(1+w_x)}
,\end{equation}
where  $\rho_{x,0}$ is the present value of $x$ components. 

However, the evolution of  DE  density with $z$ is expressed in terms of $\omega(z)$ as a solution of Eq~\eqref{eq_2} as

\begin{equation}\label{eq_7}
f_{DE}(z) \equiv \frac{\rho_{\text{DE}}(z)}{\rho_{\text{DE}}(0)} = e^{ \left[- 3 \int_0^z \frac{1 + w_{\text{DE}}(z')}{1+z'} dz' \right]}
\end{equation}
The function $f_{DE}(z)$ characterizes the evolution of the dark energy density. In the case of $\Lambda$CDM with $\omega_{\text{DE}} = -1$, this function simplifies to $f_{\text{DE}}(z) = 1$, leading to a constant normalized dark energy density parameter given by:
\begin{equation}\label{eq_6}
\tilde{\Omega}_{\text{DE}}(\equiv \rho_{DE}(z)/\rho_{critical}(0)) = \Omega_{\text{DE},0} = \text{constant}
\end{equation}

In GEDE framework, the evolution of normalized dark energy density parameter is represented as:
\begin{equation}\label{eq_6a}
\tilde{\Omega}_{DE}(z)=\rho_{DE}(z)/\rho_{critical}(0)=\Omega_{DE}(z)\times H^2(z)/H_0^2 
\end{equation}
which evolves as \citep{Li:2020ybr,Yang:2021eud}

\begin{equation}\label{eq_10}
  f_{DE}(z)\equiv \frac{\tilde{\Omega}_{\text{DE}}(z)}{\Omega_{\text{DE},0} } = \left( \frac{1 - \tanh\left(\Delta \times \log_{10} \left( \frac{1+z}{1+z_t} \right)\right)}{1 + \tanh\left(\Delta \times \log_{10} (1 + z_t)\right)} \right).
\end{equation}
The dimensionless Friedmann equation in terms of normalized density parameters for GEDE frameworks is given by
\begin{equation}\label{eq_110}
\begin{split}
    E^2(z) = & \Omega_{rad,0}(1 + z)^4 + \Omega_{m0}(1 + z)^3 \\
    & + \Omega_{\text{DE},0} \left( \frac{1 - \tanh\left(\Delta \times \log_{10} \left( \frac{1+z}{1+z_t} \right)\right)}{1 + \tanh\left(\Delta \times \log_{10} (1 + z_t)\right)} \right),
\end{split}
\end{equation}
Here, \( z_t \) represents the epoch where the matter energy density and the DE density are equal, that is, 
\(\rho_m(z_t) = \rho_{DE}(z_t) \implies \Omega_{m0} (1 + z_t)^3 = \tilde{\Omega}_{DE}(z_t)\). Hence, \( z_t \) is a derived parameter, not a free parameter. The parameter \( \Delta \) is dimensionless and free, with important characteristics. When \( \Delta \) is set to zero, Eq~\eqref{eq_10} corresponds to the standard \(\Lambda\)CDM model i.e, with constant $\omega=-1$. On the other hand, when \( \Delta \) is set to 1 and \( z_t \) is set to 0, the model recovers the PEDE model \citep{Li:2019yem}.

The equation of state of the GEDE model can be obtained from \eqref{eq_10} by using  the following expression 
\begin{equation}\label{eq_60}
\omega(z)=\frac{1}{3}\frac{d\ln \tilde{\Omega}_{DE}}{dz}(1+z)-1
\end{equation}
which gives
\begin{equation}\label{eq_601}
\omega(z)=-1-\frac{\Delta}{3\ln (10)}\left[ 1+\tanh\left( \Delta \log_{10}\left(\frac{1+z}{1+z_t}\right)\right)\right].
\end{equation}
%%%%%%%%%%%%%%%%%%%%%%%%%%%%%%%%%%%%%%%%%%%%%%%%%%%%%%%%%%%%%
\section{Methodology and Datasets}\label{sec_3}
In this analysis, we constrain the parameters of the GEDE model using \texttt{nested sampling} implemented via the \texttt{PyPolyChord} library \footnote{\url{https://github.com/PolyChord/PolyChordLite}}, which allows for efficient exploration of the high-dimensional parameter space while simultaneously computing the Bayesian evidence. In our analysis, we define uniform priors on the model parameters, $h \in [0.0, 1.0]$, $\Omega_{m0} \in [0.0,1.0]$, and $\Delta \in [-5.0,5.0]$, implemented via the \texttt{UniformPrior} class in PyPolyChord. The likelihood function compares theoretical predictions with different observational datasets. We configure PyPolyChord with $n_{\text {live }}=300$ and enable clustering to better capture multimodal posteriors. The sampler outputs posterior samples and estimates the Bayesian evidence numerically by the algorithm. Post processing and visualization of the posterior distributions and parameter correlations are performed with the \texttt{getdist} package \footnote{\url{https://github.com/cmbant/getdist}} \cite{lewis2025getdist}. We utilize multiple observational datasets to compare theoretical predictions with observations, including Baryon Acoustic Oscillations, Type Ia Supernovae, and CMB distance priors. Below, we describe each dataset in detail and explain their likelihoods.
%%%%%%%%%%%%%%%%%%%%%%%%%%%%%%%%%%%%%%%%%%%%%%%%%%%%%%%%%%
\begin{itemize}
     \item \textbf{Baryon Acoustic Oscillation :} We incorporate 13 recent Baryon Acoustic Oscillation (BAO) measurements from DESI Data Release 2 (DR2)\footnote{\url{[https://github.com/CobayaSampler/bao\_data}} \cite{DESI:2025zgx}. These measurements are derived from multiple tracers, including the Bright Galaxy Sample (BGS), Luminous Red Galaxies (LRG1, LRG2, and LRG3), Emission Line Galaxies (ELG1, and ELG2), Quasars (QSO), and Lyman-$\alpha$ forests. In this work, we use the combined LRG3+ELG1 results instead of the individual LRG3 and ELG1 measurements since they are correlated. The BAO scale is set by the sound horizon at the drag epoch ($z_d \approx 1060$), computed as: $r_d = \int_{z_d}^{\infty} \frac{c_s(z)}{H(z)} \, dz,$ where $c_s(z)$ depends on the baryon-to-photon density ratio. In flat $\Lambda$CDM, $r_d = 147.09 \pm 0.2 $ Mpc \cite{Planck:2018vyg}. We compute the Hubble distance $D_H(z) = \frac{c}{H(z)}$, the comoving angular diameter distance $D_M(z) = c \int_0^z \frac{dz'}{H(z')}$, and the volume-averaged distance: $D_V(z) = \left[ z D_M^2(z) D_H(z) \right]^{1/3}$ \citep{Hogg:1999ad}. Model constraints are derived from ratios such as $D_M/r_d$, $D_H/r_d$, $D_V/r_d$, and $D_M/D_H$. The BAO chi-squared is defined as: $\chi^2_{\mathrm{BAO}} = \Delta \mathbf{D}^\intercal \mathbf{C}^{-1} \Delta \mathbf{D},$ where $\Delta \mathbf{D} = \mathbf{D}_{\mathrm{obs}} - \mathbf{D}_{\mathrm{th}}$ and $\mathbf{C}^{-1}$ is the inverse covariance matrix.
     
     \item \textbf{Type Ia supernova :} We also use three different SNe Ia samples to enhance the constraints on cosmological parameters. The first is the PantheonPlus (PP) sample\footnote{\url{https://github.com/PantheonPlusSH0ES/DataRelease}} \cite{brout2022pantheon}, which contains 1,590 light curves from 1,550 SNe Ia covering the redshift range $0.01 \leq z \leq 2.26$. To reduce systematic effects from peculiar velocities, we exclude light curves with $z < 0.01$. The second dataset is the DES-SN5Y sample\footnote{\url{https://github.com/des-science/DES-SN5YR}} \cite{abbott2024dark}, which includes 1,829 SNe Ia in the range $0.025 \leq z \leq 1.12$. This set consists of 1,635 SNe observed directly by DES, along with 194 nearby ($z < 0.1$) SNe Ia from the CfA/CSP Foundation sample. Finally, we use the Union3 compilation \cite{rubin2025union}, which provides 2,087 cosmologically useful SNe Ia drawn from 24 datasets, spanning the redshift range $0.050 \leq z \leq 2.26$ In our analysis, we marginalize over $\mathcal{M}$ parameter; for further details, see Equations (A9–A12) of \cite{goliath2001supernovae}.
     
    \item \textbf{CMB distance priors} Finally, we also make use of the CMB distance priors summarized in \cite{Chen:2018dbv}. In our analysis, we use three quantities: the acoustic scale $\ell_A$, the shift parameter $R$, and the physical baryon density $\Omega_b h^2$. The acoustic scale $\ell_A$ characterizes the transverse features of the CMB temperature power spectrum, determining the spacing between acoustic peaks. The shift parameter $R$ encodes the projection of the sound horizon along the line of sight and therefore affects the relative heights of the peaks. They are defined as $\ell_A = (1 + z_*) \frac{\pi D_A(z_*)}{r_d}$, $R(z_*) = \sqrt{\Omega_m H_0^2} \, \frac{(1 + z_*) D_A(z_*)}{c},$ where $D_A(z_*)$ is the angular diameter distance to the redshift of decoupling $z_*$ and $r_d$ is the comoving sound horizon at the drag epoch. The covariance matrix for $(\ell_A, R, \Omega_b h^2)$ is provided in Table~I of Ref.~\cite{Chen:2018dbv}.
\end{itemize}
%%%%%%%%%%%%%%%%%%%%%%%%%%%%%%%%%%%%%%%%%%%%%%%%%%%%%%%%%%%%

The posterior distributions of the GEDE model parameters are derived by maximizing the total likelihood function, expressed as: $\mathcal{L}_{\text{tot}} = \mathcal{L}_{\text{BAO}} \times \mathcal{L}_{\text{SNe Ia}} \times \mathcal{L}_{\text{CMB}}.$ In this analysis, we consider the parameters \( h = H_0/100 \), \( \Omega_{m0} \), and \( \Delta \) as free parameters. The present-day radiation density parameter, \( \Omega_{rad,0} \), is computed via the relation $\Omega_{\mathrm{rad}} = 2.469 \times 10^{-5}\, h^{-2} \left( 1 + 0.2271\, N_{\mathrm{eff}} \right)$~\cite{komatsu2009five}, 
where $N_{\mathrm{eff}} = 3.04$ is the standard effective number of relativistic species~\cite{mangano2002precision}. Accordingly, the DE density parameter today is obtained from the flatness condition: $\Omega_{DE,0} = 1 - \Omega_{m0} - \Omega_{rad,0}.$ As a result, both \( \Omega_{rad,0} \) and \( \Omega_{DE,0} \) are not treated as independent parameters, since they are fully determined by the remaining ones. We also compute the Bayes factor \cite{Trotta:2008qt}, defined as $B_{i0} = \frac{p(d|M_i)}{p(d|M_0)}$, where $p(d|M_i)$ and $p(d|M_0)$ represent the Bayesian evidences for the GEDE model ($M_i$) and the reference model $M_0$, respectively. In our analysis, $M_0$ corresponds to $\Lambda$CDM model, while $M_i$ represents the GEDE model. Since analytical calculation of the Bayesian evidence is difficult, we use PolyChord, which computes it numerically through a nested sampling algorithm. We report the natural logarithm of the Bayes factor, $|\ln(B_{i0})|$, and interpret the results using Jeffreys' scale \cite{jeffreys1961theory}: $|\ln(B_{i0})| < 1$ indicates inconclusive evidence; $1 \leq |\ln(B_{i0})| < 2.5$ suggests weak support for the GEDE model; $2.5 \leq |\ln(B_{i0})| < 5$ implies moderate support; and $|\ln(B_{i0})| > 5$ represents strong support for the GEDE model. Negative values of $\ln(B_{i0})$ indicate a preference for the reference $\Lambda$CDM model.
%%%%%%%%%%%%%%%%%%%%%%%%%%%%%%%%%%%%%%%%%%%%%%%%%%%%%%%%%%%%%
\begin{figure}
\centering
\includegraphics[scale=0.55]{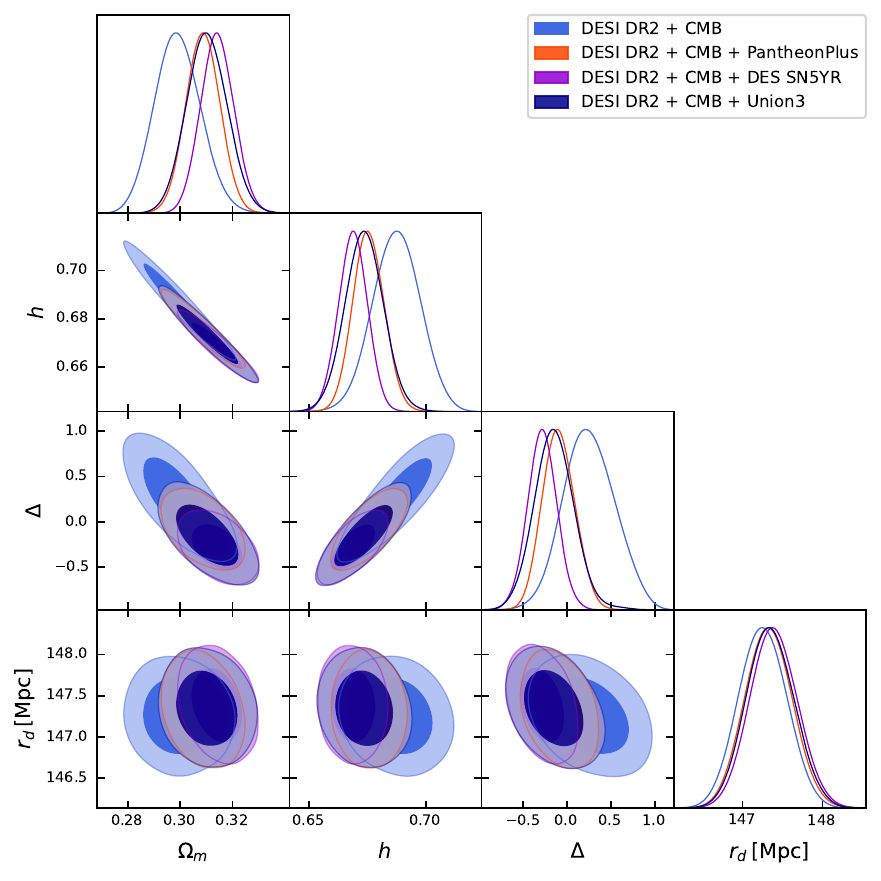}
\caption{The figure shows the contour plot of the GEDE model showing the 68\% (1$\sigma$) and 95\% (2$\sigma$) confidence levels, using DESI DR2, CMB, and Type Ia Supernova datasets (PP, DES-SN5Y, and Union3).}\label{fig_1}
\end{figure}
%%%%%%%%%%%%%%%%%%%%%%%%%%%%%%%%%%%%%%%%%%%%%%%%%%%%%%%%%%%%%
\begin{figure}
\centering
\includegraphics[scale=0.55]{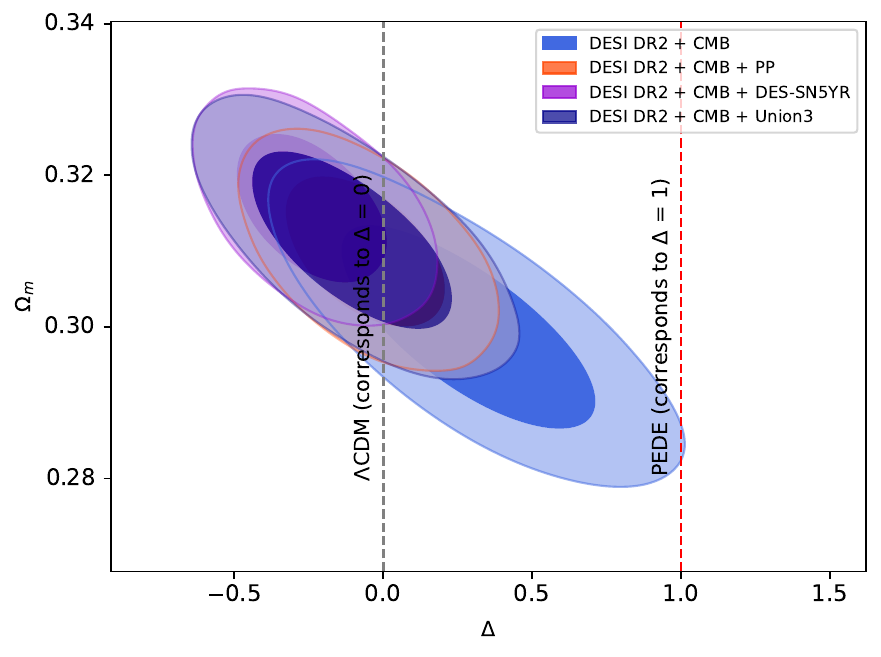}
\caption{The figure shows the marginalized posterior distributions in the $\Delta$–$\Omega_{m0}$ plane for the GEDE model at 68\% (1$\sigma$) and 95\% (2$\sigma$) confidence levels, using DESI DR2, CMB, and Type Ia Supernova datasets (PP, DES-SN5Y, and Union3). The vertical dashed gray line at $\Delta = 0$ represents the standard $\Lambda$CDM limit, while the vertical dashed gray line at $\Delta = 1$ corresponds to the PEDE scenario.}\label{fig_2}
\end{figure}
%%%%%%%%%%%%%%%%%%%%%%%%%%%%%%%%%%%%%%%%%%%%%%%%%%%%%%%%%%%
\begin{table*}
\begin{tabular}{llcc}
\hline
\textbf{Parameter} & \textbf{Dataset} & \textbf{$\Lambda$CDM} & \textbf{GEDE} \\ 
\hline
& CMB + DESI DR2 & $0.684 \pm 0.004$ & $0.689 \pm 0.010$ \\
$h$ & CMB + DESI DR2 + PP & $0.683 \pm 0.004$ & $0.675 \pm 0.006$ \\
& CMB + DESI DR2 + DES-SN5Y & $0.681 \pm 0.004$ & $0.668 \pm 0.006$ \\
& CMB + DESI DR2 + Union3 & $0.683 \pm 0.004$ & $0.674 \pm 0.008$ \\
\hline
& CMB + DESI DR2 & $0.303 \pm 0.005$ & $0.299 \pm 0.008$ \\
$\Omega_{mo}$ & CMB + DESI DR2 + PP & $0.305 \pm 0.007$ & $0.309 \pm 0.006$ \\
& CMB + DESI DR2 + DES-SN5Y & $0.308 \pm 0.005$ & $0.315 \pm 0.006$ \\
& CMB + DESI DR2 + Union3 & $0.305 \pm 0.005$ & $0.311 \pm 0.007$ \\
\hline
& CMB + DESI DR2 & --- & $0.30{\pm 0.28}$ \\
$\Delta$ & CMB + DESI DR2 + PP & --- & $-0.06 \pm 0.18$ \\
& CMB + DESI DR2 + DES-SN5Y & --- & $-0.24{\pm 0.17}$ \\
& CMB + DESI DR2 + Union3 & --- & $-0.10 \pm 0.22$ \\
\hline
& CMB + DESI DR2 & $147.24 \pm 0.28$ & $147.25 \pm 0.30$ \\
$r_d$ (Mpc) & CMB + DESI DR2 + PP & $147.18 \pm 0.28$ & $147.33 \pm 0.29$ \\
& CMB + DESI DR2 + DES-SN5Y & $147.12 \pm 0.28$ & $147.39 \pm 0.29$ \\
& CMB + DESI DR2 + Union3 & $147.18 \pm 0.28$ & $147.35 \pm 0.30$ \\
\hline
& CMB + DESI DR2 & 0 & 1.96 \\
$\lvert \ln(B_{\Lambda \mathrm{CDM}, \, \mathrm{Model}}) \rvert$
 & CMB + DESI DR2 + PP & 0 & 2.65 \\
& CMB + DESI DR2 + DES-SN5Y & 0 & 1.44 \\
& CMB + DESI DR2 + Union3 & 0 & 2.34 \\
\hline
\end{tabular}
\caption{This table presents the numerical values obtained for the $\Lambda$CDM and GEDE models at the 68\% (1$\sigma$) confidence level, using different combinations of DESI DR2 BAO datasets with the CMB and various SNe Ia samples.}
\label{tab_1}
\end{table*}

%%%%%%%%%%%%%%%%%%%%%%%%%%%%%%%%%%%%%%%%%%%%%%%%%%%%%%%%%
\begin{figure*}
\centering
\includegraphics[width=\linewidth]{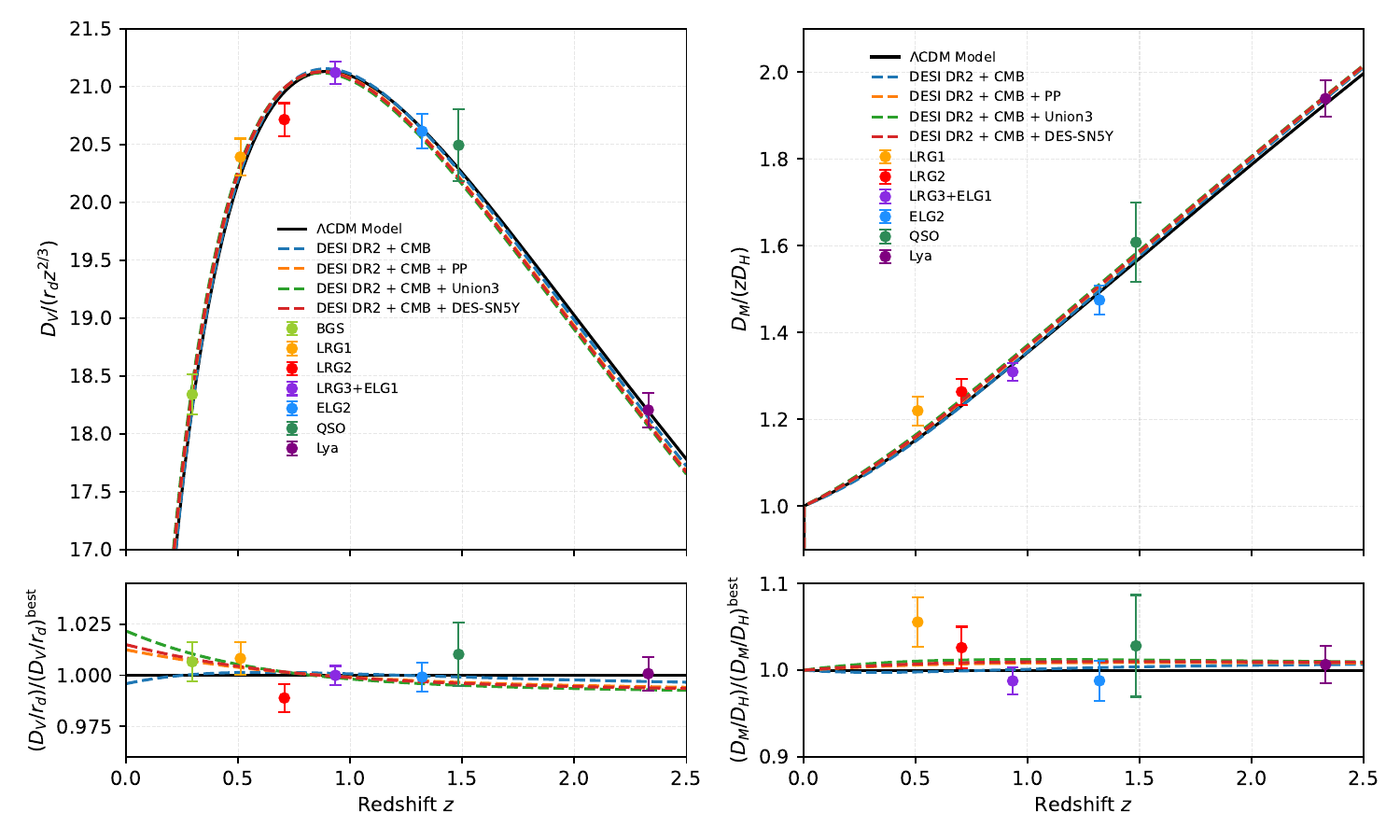} 
\caption{The figure shows the evolution of the angle-averaged distance, $D_V(z)/r_d$, and the ratio of transverse to line-of-sight comoving distances, $D_M(z)/D_H(z)$, scaled by $z^{-2/3}$ and $z^{-1}$, respectively, in the upper panel. The bottom panel displays the corresponding residuals.}\label{fig_3}
\end{figure*}
%%%%%%%%%%%%%%%%%%%%%%%%%%%%%%%%%%%%%%%%%%%%%%%%%%%%%%%%%%%%%
\begin{figure*}
\centering
\includegraphics[width=1.0\textwidth]{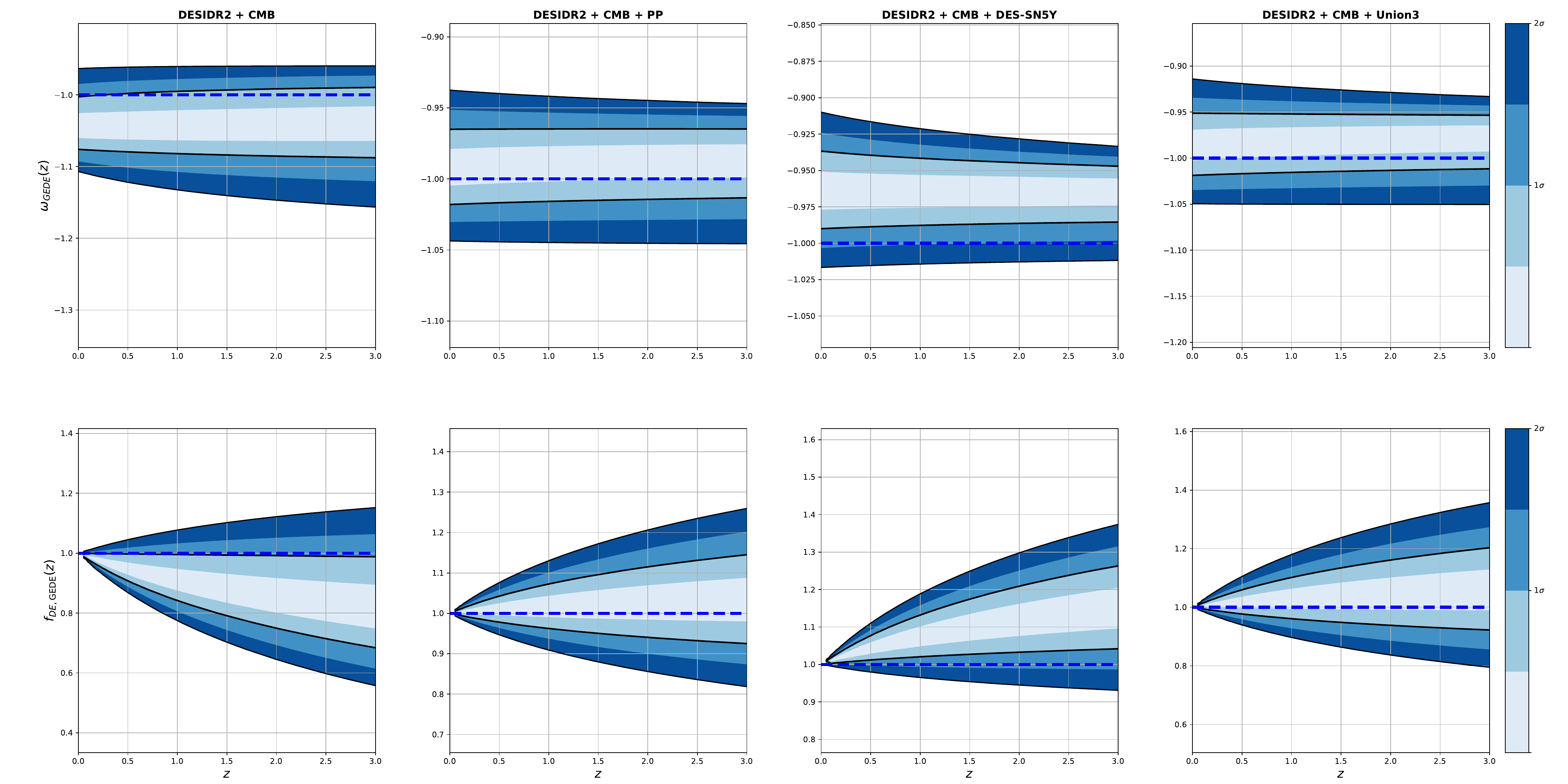}
\caption{Marginalized constraints on the dark energy equation of state, $\omega(z)$, and normalized energy density, $f_{\mathrm{DE}}(z)$, within the framework of the GEDE model. }\label{fig_4}
\end{figure*}
%%%%%%%%%%%%%%%%%%%%%%%%%%%%%%%%%%%%%%%%%%%%%%%%%%%%%%%%%%%%%%
\section{Results}\label{sec_4}
Fig.~\ref{fig_1} presents the contour plot of the GEDE model parameters at 68\% (1$\sigma$) and 95\% (2$\sigma$) confidence levels, derived from various combinations of cosmological datasets, including DESI DR2, CMB, and multiple supernova compilations (PP, DES-SN5Y, and Union3). The diagonal panels display the 1D marginalized posterior distributions for each cosmological parameter such as $h$, $\Omega_{m0}$, and $\Delta$ indicating their most probable values and associated uncertainties. The off diagonal panels show the 2D joint posterior distributions for parameter pairs, with the inner and outer contours representing the 68\% (1$\sigma$) and 95\% (2$\sigma$) confidence levels, respectively.

Table~\ref{tab_1} shows the mean values and 1$\sigma$ uncertainties for each parameter in the $\Lambda$CDM and GEDE models. Focusing on the Hubble constant $h$, the $\Lambda$CDM model predicts values in the range $0.681 - 0.684$, consistent with the Planck 2018 estimate of $h = 0.674 \pm 0.005$ \citep{Planck:2018vyg}. The GEDE model remains broadly compatible with Planck, with deviations for each dataset combination as follows: CMB + DESI DR2 gives $h = 0.689 \pm 0.010$ ($1.34\sigma$ deviation), CMB + DESI DR2 + PP yields $h = 0.675 \pm 0.006$ ($0.13\sigma$ deviation), CMB + DESI DR2 + DES-SN5Y gives $h = 0.668 \pm 0.006$ ($0.77\sigma$ deviation), and CMB + DESI DR2 + Union3 gives $h = 0.674 \pm 0.008$ (exactly consistent, $0.00\sigma$ deviation). All deviations are smaller than $1.4\sigma$, indicating that the GEDE model is fully compatible with the Planck 2018 measurement of the Hubble constant, with only minor variations depending on the dataset combination.

We also observe that the matter density parameter $\Omega_{m0}$ predicted by the GEDE model is consistent with the value reported by the Planck collaboration ($\Omega_{m0} = 0.315 \pm 0.007$). Specifically, the GEDE model yields $\Omega_{m0} = 0.299 \pm 0.008$ for CMB + DESI DR2 ($1.51\sigma$ deviation from Planck), $\Omega_{m0} = 0.309 \pm 0.006$ for CMB + DESI DR2 + PP ($0.65\sigma$ deviation), $\Omega_{m0} = 0.315 \pm 0.006$ for CMB + DESI DR2 + DES-SN5Y (exactly consistent, $0.00\sigma$ deviation), and $\Omega_{m0} = 0.311 \pm 0.007$ for CMB + DESI DR2 + Union3 ($0.40\sigma$ deviation). All these values lie within a reasonable range of the Planck estimate, with deviations smaller than $1.6\sigma$, indicating that the GEDE model remains broadly consistent with current cosmological observations.

Fig.~\ref{fig_2} presents the marginalized posterior distributions in the $\Delta$–$\Omega_{m0}$ plane. It can be observed that in the GEDE model, when we combine DESI DR2 with CMB, the GEDE prefers positive values of $\Delta$, whereas when different SNe Ia calibrations are added to DESI DR2 and CMB, the GEDE prefers negative values of $\Delta$, which corresponds to the injection of DE at earlier redshifts. \cite{lodha2025desi,lodha2025extended} also predict similar values when they tested the GEDE model in light of DESI DR1 and DESI DR2, respectively. In \citep{Yang:2021eud}, these negative values were subsequently tested against the matter power spectrum, which further reinforced the viability of the GEDE framework.

Fig.~\ref{fig_3} shows the evolution of the angle-averaged distance, $D_V(z)/r_d$, and the ratio of transverse to line-of-sight comoving distances, $D_M(z)/D_H(z)$, scaled by $z^{-2/3}$ and $z^{-1}$, for the GEDE model, using the numerical values obtained from different combinations of the datasets. These evolutions are then compared against the $\Lambda$CDM predictions and the different tracers, along with their associated error bars. In the upper panel, it can be observed that in both cases, the GEDE model predictions show close agreement with the $\Lambda$CDM predictions at $z < 1.0$. Deviations appear at $z > 1.0$, but these differences remain relatively small. In the bottom-left panel, which shows the residual plots, it can be seen that the numerical evaluation from DESI DR2 + CMB shows close agreement with the $\Lambda$CDM model. However, when different SNe Ia calibrations are added, the GEDE model exhibits deviations from the $\Lambda$CDM predictions. These behaviors can be understood in terms of the value of $\Delta$: in light of the DESI DR2 + CMB data, the preferred value of $\Delta$ in the GEDE model is close to that of the $\Lambda$CDM model, but when different SNe Ia calibrations are added, noticeable deviations from $\Lambda$CDM become apparent. In the bottom right panel, it can be observed that the GEDE model shows close agreement with the $\Lambda$CDM model. Furthermore, the $\Lambda$CDM and GEDE models predict values of $r_d$ that are consistent with those reported by Planck 2018 \cite{Planck:2018vyg}. When comparing the predicted $r_d$ values from the $\Lambda$CDM model to those from the GEDE model, we find that the GEDE model exhibits deviations of 0.023$\sigma$ for the CMB + DESI DR2, 0.12$\sigma$ for CMB + DESI DR2 + Pantheon$^+$, 0.27$\sigma$ for CMB + DESI DR2 + DES-SN5Y, and 0.24$\sigma$ for CMB + DESI DR2 + Union3.

Fig.~\ref{fig_2} presents the marginalized posterior distributions in the $\Delta$–$\Omega_{m0}$ plane. It can be observed that in the GEDE model, when we combine DESI DR2 with CMB, the GEDE prefers positive values of $\Delta$, whereas when different SNe Ia calibrations are added to DESI DR2 and CMB, the GEDE prefers negative values of $\Delta$, which corresponds to the injection of DE at earlier redshifts. \cite{lodha2025desi,lodha2025extended} also predict similar values when they tested the GEDE model in light of DESI DR1 and DESI DR2, respectively. In \citep{Yang:2021eud}, these negative values were subsequently tested against the matter power spectrum, which further reinforced the viability of the GEDE framework.

Fig.~\ref{fig_4} presents the Marginalized constraints on the dark energy EoS $\omega(z)$ and the normalized DE density evolution $f_{\mathrm{DE}}(z)$ within the GEDE framework, shown with 1$\sigma$ and 2$\sigma$ confidence bands, derived from various combinations of cosmological datasets. The GEDE model shows a sharp emergence but then gradually approaches $\omega = -1$ without ever crossing it. The present-day values of the EoS reflect this behavior: for CMB+DESI DR2 we obtain $\omega(0) = -1.041$, while the inclusion of supernova data shifts the results closer to $-1$, with $\omega(0) = -0.991$ for CMB+DESI DR2+PP, $\omega(0) = -0.964$ for CMB+DESI DR2+DES-SN5Y, and $\omega(0) = -0.985$ for CMB+DESI DR2+Union3. These results indicate that GEDE is closest to $\Lambda$CDM when only CMB+DESI DR2 data are used, while the inclusion of different SNe Ia calibrations shifts the preference toward $\omega > -1$. The corresponding $f_{\mathrm{DE}}(z)$ panels show that while the DE density may evolve differently from $\Lambda$CDM at higher redshifts, all cases converge to $f_{\mathrm{DE}}(0) = 1$ at the present epoch.

The Bayesian comparison is used here to contrast the standard flat $\Lambda$CDM model against the GEDE model, based on the natural logarithm of the Bayes factor $\ln(BF)$ computed across various dataset combinations. When only DESI DR2 and CMB data are used, the result $\ln(BF) = 1.96$ corresponds to weak support for the GEDE model, indicating that both models remain broadly comparable. When supernova datasets are included, the preference for the GEDE model becomes clearer. For the combination CMB + DESI DR2 + PP, a value of $\ln(BF) = 2.65$ is obtained, which represents moderate support for GEDE. Similarly, the Union3 and DES-SN5Y combinations yield $\ln(BF) = 2.34$ and $\ln(BF) = 1.44$, corresponding to weak to moderate and weak support, respectively. Overall, while both models are consistent with the observational data, the inclusion of supernova samples tends to shift the evidence in favor of the GEDE model, particularly for the PP and Union3 datasets.

\section{Conclusion}\label{sec_5}
In this work, we have tested the GEDE model against the latest cosmological datasets, including DESI DR2 BAO measurements, CMB data, and three different Type Ia supernova samples (Pantheon$^+$, DES-SN5Y, and Union3). Using \texttt{PyPolyChord} for nested sampling, we constrained the key cosmological parameters and computed the Bayesian evidence to compare GEDE with the standard $\Lambda$CDM paradigm. Our analysis shows that GEDE yields values of the Hubble constant $h$ and matter density $\Omega_{m0}$ that are fully consistent with the Planck 2018 results, with deviations always below $1.6\sigma$. The parameter $\Delta$, which governs deviations from $\Lambda$CDM, is found to be positive when using CMB+DESI DR2 alone but turns negative when supernova datasets are added, implying an earlier injection of DE around $z_t \simeq 0.5$. The predicted value of $r_d$ from the $\Lambda$CDM model deviates from the GEDE predicted value by within 0.3$\sigma$, depending on the combination of datasets used.

The reconstructed EoS $\omega(z)$ and DE density $f_{\mathrm{DE}}(z)$ reveal that GEDE exhibits phantom-like behavior ($\omega < -1$) with CMB+DESI DR2, but shifts toward quintessence-like behavior ($\omega > -1$) once supernova data are included. GEDE shows its largest deviation from $\Lambda$CDM around $z \sim 1$, where it provides a slightly better fit to the data, while at low redshifts ($z \lesssim 0.5$) the model is driven back toward $\omega \approx -1$, reducing its advantage. The Bayesian comparison indicates that CMB+DESI DR2 alone provides only weak support for GEDE ($\ln(BF)=1.96$). When supernova datasets are included, the evidence in favor of GEDE becomes clearer, with moderate support for PP ($\ln(BF)=2.65$), weak-to-moderate support for Union3 ($\ln(BF)=2.34$), and weak support for DES-SN5Y ($\ln(BF)=1.44$). Overall, GEDE remains a viable extension of $\Lambda$CDM, consistent with current observations and mildly favored when supernova datasets are considered. However, its advantage is modest, as the model tends to “mellow” toward $\Lambda$CDM at low redshift. Future high-precision measurements of the expansion history, particularly at $z \sim 1$, will be crucial in determining whether the emergent features of GEDE represent a genuine departure from $\Lambda$CDM or simply statistical fluctuations in current data.

\bibliographystyle{elsarticle-num}
\bibliography{mybib}

\end{document}